# Strain self-accommodation during growth of 14H type long-period stacking ordered (LPSO) structures in Mg-Zn-Gd alloy

Xin-Fu Gu[a,b]*, Tadashi Furuhara[b], Takanori Kiguchi[b], Toyohiko J. Konno[b], Leng Chen[a], Ping Yang[a]

[a] School of Materials Science and Engineering, University of Science and Technology Beijing, Beijing, 100083, China

[b] Institute for Materials Research, Tohoku University, Sendai, 980-8577, Japan

*Corresponding author: Xin-Fu Gu, xinfugu@gmail.com

**ABSTRACT**

Cooperative growth of the structure units (SUs) in a 14H type long-period stacking ordered (LPSO) structure has been observed, and there is no obvious accumulation of transformation strain at the growth fronts. The atomic structures at this front are further characterized by Cs-corrected scanning transmission electron microscopy, and the partial dislocations associated SUs are uniquely defined based on the observations at different zone axes. It is found that the Burgers vectors of neighbouring partials are alternatively opposed so that the transformation strain is self-accommodated. Furthermore, this self-accommodation mechanism is rationalized by the elastic interaction energy for combinations of different partials.

Long-period stacking ordered (LPSO) structures are known as an important strengthening phase in Mg-M-RE based alloys (M: Zn, Cu, Ni, Al, or Co, RE: Y, Gd, Tb, Dy, Ho, Er, or Tm) [1-6]. The LPSO structure can be treated as lamellar structure with alternative stacking of FCC structural units (SUs) and Mg layers (HCP structure) on $(0001)_{hcp}$ plane, and the SUs also enrich with solute elements M and RE [7-9]. The SUs in the commonly observed LPSO structure 10H, 18R, 14H, and 24R are separated by 1, 2, 3 and 4 of Mg layers, respectively [10, 11], where the Ramsdell's notation indicates the total number of $(0001)_{hcp}$ layers in the hexagonal unit cell and the followed letter H or R to specify the lattice type. Accordingly, there are two SUs in a unit cell of H type LPSO structures, while three SUs are in a unit cell of R type LPSO structures.

The formation of a LPSO structure involves a change in both structure and composition [12-15]. Specifically, the FCC SU is transformed from HCP structure by operating of a $<1\bar{1}00>_{hcp}/3$ type Shockley partial dislocation on the basal plane. Meanwhile, the SU is synchronized with M and RE elements during the transformation. This kind of phase transformation is so-called displacive-diffusional transformation [15-18]. It has been found that the LPSO structure grows by a ledge mechanism observed by atomic resolved high angle annular dark field (HAADF)-scanning transmission electron microscopy (STEM) [12]. The growth ledge is associated with a Shockley partial dislocation, and this defect is generally termed as disconnection [12, 19]. Since the transformation product is constrained by surrounding matrix, the transformation strain would cause elastic distortion in the matrix with the generation and movement of the disconnections. As for the transformation from HCP structure to the four-layer-height SU, the shear strain caused by the disconnection is about 0.1, which is significant large. Therefore, it is necessary to consider the strain accommodation mechanism during the precipitation of LPSO structures [17, 20].

The transformation strain can be accommodated in terms of diffusion, plastic deformation in matrix or self-accommodation by twin or slip in the product phase etc. [16, 17]. Since the interface between a SU and Mg matrix is $(0001)_{hcp}$ coherent interface, the transformation would be possibly relaxed by a long-range volume diffusion, and the high-speed path for diffusional accommodation would be impossible due to high coherency in the interface. Thereby, the self-accommodation mechanism may be dominant. Periodic arrangements of the SUs in the LPSO structure can be generated by periodic operating of the Shockley partial dislocations. However, there are three equivalent Burgers vector for a Shockley partial in the basal plane, thus there are several possible combinations of partial variants for neighbouring SUs. As a result, the overall transformation strain would vary with the possible combinations of the partials.

Zhu et al. [12] proposed that the preferred combination would be the case that the summation of possible partial dislocations associated the SUs to be zero. Similar configurations are observed in some simple FCC/ HCP systems, such as the partial dislocation configuration in the interface in a Al-Ag alloy [21, 22], Mg-Sn alloy [23] etc.. In addition, we have theoretically examined the possible configurations of partials for the LPSO structures based on the elastic energy calculation, and drew similar conclusion that the energy will reach their minimum when the macroscopic net strain around growth tip is minimized [20]. The most preferred combination of partials for H type LPSO structure is that the shears for neighboring SUs should be opposite, while the preferred combination for R type LPSO structure is that three alternative partials are operated. Nevertheless, the actual dislocation configuration at growth interface in the LPSO structure has not been experimentally clarified, despite of their importance in Mg-M-RE based alloys. The understanding of the transformation strain accommodation mechanism would also be benefit to the possible control of the LPSO structure by external strain field, such as pre-deformation. In this study, we aim to clarify the possible dislocation configuration at the transformation front between the LPSO structure and Mg matrix by transmission electron microscopy, and the possible configurations will be discussed in view of elastic interaction energy.

The as-casted $Mg_{97}Zn_1Gd_2$ (at. %, default) alloy was solution treated at 520°C for 2h, and then aged at 500°C for 4h. The microstructure is characterized by SEM (JOEL 7001F), Cs-corrected STEM (Titan$^3$ G2 60-300), and 3-D atom probe (CAMECA Leap 4000). The procedures to prepare the sample and the conditions for microstructure observation are the same as our previous work [24].

Fig. 1 shows the microstructure of the aged sample at 520°C for 4h. The microstructure for as-casted and solution treated sample can be found in Fig. S1 in the supplementary materials. As shown by the SEM image in Fig. 1(a), plate-like precipitates are precipitated during ageing process, and these precipitates have a LPSO structure according to previous work [25]. Fig. 1(b) shows a low-mag HAADF-STEM image viewed along $[1\bar{1}00]_{hcp}$ direction. According to the Z contrast principle in the HADDF-STEM image [26], the bright contrast in the image corresponds to the precipitates due to enrichment of solute atoms Gd and/or Zn, since the atomic number (Z) for Mg, Zn and Gd is 12, 30 and 64, respectively. The tips (i.e. transformation/growth front) of the LPSO structure and the growth ledge are observed as indicated by the arrows. The LPSO structure is thickening by the ledge mechanism in agreement with the previous work [12]. Interestingly, the bright linear contrasts, i.e. the SUs in the LPSO structure, align well at the growth front, and it implies that the lengthening of the SUs may cooperatively proceed. The cooperative growth phenomena are common for different LPSO precipitates in Fig. 1(a) and also at different ageing time. According to the enlarged view of the growth front enclosed with the yellow rectangle in Fig. 1(b), the number of cooperative SUs from top to bottom is 4, 2 and 6, which are all even numbers, respectively. Fig. 1(c) shows a high magnification of the precipitates. The precipitate shows characteristic features of LPSO structure. It consists of fcc SUs indicated by yellow line segments and the SUs are separated by 3 Mg layers on (0001) plane. Therefore, the precipitate is 14H type

LPSO structure. The images viewed at $[1\bar{2}10]_{hcp}$ and $[1\bar{1}00]_{hcp}$ in Fig. 1(c) also show that the LPSO structure is not well-ordered. The element mapping of the LPSO structure by 3DAP is shown in Fig. 1(d), and the SU is synchronized with Zn and Gd elements. The composition of the LPSO structure is determined to be $Mg_{89.1}Gd_{5.8}Zn_{5.1}$ and lower than the well-ordered structure $Mg_{83.4}Gd_{9.5}Zn_{7.1}$ [7], in agreement with the observation from Fig. 1(c).

A schematic diagram of 14H type LPSO structure is shown in Fig. 2(a) viewed along $[11\bar{2}0]$ direction. The stacking sequence of 14H type LPSO along the close-packed planes is AB*ABCA*CAC*ACBA*B (Bold italic letters indicate the stacking sequence in the SU). The four layer height SUs is also highlighted by grey box in Fig. 2(a). The transformation of HCP→FCC SU is shown in Fig. 2(b). The ABAB stacked HCP structure is changed to ABCA stacking sequence in the SU by operating a Shockley partial. Three equivalent partials ($s_1$, $s_2$ and $s_3$) for this change are shown in Fig. 2(c). In addition, the partial to obtain the neighboring SU unit from HCP is -$s_1$, -$s_2$ or -$s_3$. Therefore, the partials for the transformation are not fixed, and we need to determine the Burgers vector of the partial dislocation associated with each SU at the growth front in order to understand the transformation mechanism.

Fig. 3 shows the atomic-resolved HAADF-STEM image used to determine the Burgers vector. Four SUs in a 14H type LPSO structure are shown in Fig. 3(a) at the zone axis of $[1\bar{2}10]_{hcp}$. The core of the partial dislocations could be identified where the transition between different stacking sequences begins. The Burgers vector for each partial dislocation associated with the SU could be determined by plotting the Burgers circuit around the dislocations. Another example can be found in the Fig. S2 in the supplementary materials. The closure failures of the Burgers circuits in Fig. 3(a) are indicated by the arrows in the figure and the failure distances are same but with opposite directions, i.e. along $[1\bar{1}00]_{hcp}/6$ or $[\bar{1}100]_{hcp}/6$. Apparently, this displacement vector is possibly the projected component of the other two partial variants $\pm[0\bar{1}10]_{hcp}/3$ or $\pm[\bar{1}010]_{hcp}/3$ along $[1\bar{2}10]_{hcp}$. Therefore, the Burgers vectors of the partials could not be identified by this single view in Fig. 3(a). There are two possible configurations as shown in Fig. 3(c). At each configuration, the partials for neighboring SUs in 14H type LPSO structure are specified. At the view direction specified, the projection of $-\mathbf{s}_2$ and $\mathbf{s}_2$ pair is the same with $-\mathbf{s}_2$ and $\mathbf{s}_3$ pair, and the partial components in both cases are same and opposite in sign. In order to fix the Burgers vector, another view of the same growth front is needed. Fig. 3(b) shows the same area at the zone axis of $[1\bar{1}00]_{hcp}$. The zone axes between Figs. 3(a) and (b) differ 30 degree from each other. By performing the Burgers circuit analysis, all of the closure failures are found to be 0. Therefore, the Burgers vectors of the partial dislocations in Fig. 3(b) should be parallel to the zone axis, and the Burgers vectors of these partials associated with the SU is determined to be $\pm[1\bar{1}00]_{hcp}/3$. The contrast of the growth front indicates that it is in an edge-on orientation, thus the growth front has a pure screw type dislocation. Therefore, the possible combination of shears in Fig. 3(c) can be discriminated. The view direction for Fig. 3(b) is shown in Fig. 3(d), and the configuration of partial dislocations as specified in pair (1) is most possible, i.e. the dislocations for neighboring SUs have opposite sign and grow in pair. In addition, no obvious distortion of atomic positions is observed around the growth front, thus the transformation strain caused by the SUs is self-accommodated by opposite shear variants. According to contrast at the growth front Fig. 3(a-b), it seems that there is no enrichment of Zn and Gd at the partial dislocations, and this is probably due to screw nature of the partials and the high ageing temperature.

The elastic interaction energy between SUs is further evaluated in order to understand the preference in the combination of different partials or shear variants. The numerical calculation procedure is the same as our previous work [20]. In this work, a dilatational strain normal to $(0001)_{hcp}$ plane is also considered, which may be caused by enrichment of solute atoms in the SU [27]. According to the observation, this strain could be over 10%. Suppose the SU has a cuboidal shape, Fig. 4(a) shows the variation of interaction energy with the spacing between two cuboidals with a *c/a* ratio as 0.1, where *a* and *c* define the size of the cuboidal as indicated in Fig. 4(a). The possible combinations of shears for two neighboring SUs in 14H type LPSO are shown in Fig. 4(b). According to Fig. 4(a), the combination of opposite shears, i.e. pair 1, has negative interaction energy, while the other two pairs have positive interaction energy. This tendency is similar to the cases without dilatational strain [20]. Negative value means that the total energy would decrease due to the elastic interaction. Comparably, the positive interaction energy can reach their maximum when the shear is at the same direction as the case shown by the dashed line in Fig. 4(a). Therefore, the neighboring SUs with two opposite shears is energetically preferred than the other two pairs, though the separation distance between SUs in 14H type LPSO cannot be explained by the elastic interaction, which may be due to the chemical interaction between solute atoms and SUs [14]. The interaction between multiple SUs is shown in Fig. 4(c), and the total energy can be further reduced due to negative interaction energy for multiple interactions between SUs. This may explain the observation in Fig. 1 that the SUs cooperative grows in multiple pairs. The low energy configuration is schematically shown in Fig. 4(d). The shear component of neighbor SUs should be opposite to each other in order to decrease total elastic energy, which is consistent with our experimental result. Similar accommodation would be expected in 18R LPSO structure which is also commonly observed. As shown in Fig. S3 in the supplementary materials, there are several possible combinations of partials for neighboring SUs, among which the neighboring SUs with the same partials will cause a large shape change or high elastic interaction energy when it is transformed from Fig. S3(a) to S3(b), and this combination is most unfavorable. The preferred combination is shown in Fig. S3(c), different partials are assigned to three successive SUs, and the summation of these three partials are zero, thus the net shape change is minimized. Therefore, the shear strains for the SUs in LPSO structure are self-accommodated to reduce the overall transformation strain and the SUs in the tip will move cooperatively during the lengthening of LPSO phases.

In summary, the accommodation mechanism of transformation strain in 14H type LPSO structure has been investigated by Cs-corrected STEM in $Mg_{97}Zn_1Gd_2$ alloy.

Cooperative growth of the SUs is found at the transformation front between the LPSO structure and Mg matrix. The dislocation for each SU at this front has been uniquely defined as $<1\bar{1}00>_{hcp}/3$ type screw dislocation. It is found that neighbouring SUs have opposite partial dislocations which effectively accommodate the shear strains caused from the transformation from HCP to FCC. The elastic interaction energy between the SUs was evaluated for different combinations of partials, and the elastic interaction energy is minimized when neighbouring SUs exhibit partials, being consistent with the observation.

**Acknowledgement**

This work was supported by the Ministry of Education, Culture, Sports, Science and Technology, Japan [Grant No.23109001] and Beijing Municipal Natural Science Foundation, China [Grant No. 2192035]. The Mg-Zn-Gd sample studied here was supplied by Kumamoto University. The authors also thank to Mr. Y. Hayasaka (the Electron Microscopy Center, Tohoku University) for his assistant during TEM work.

**Figures**

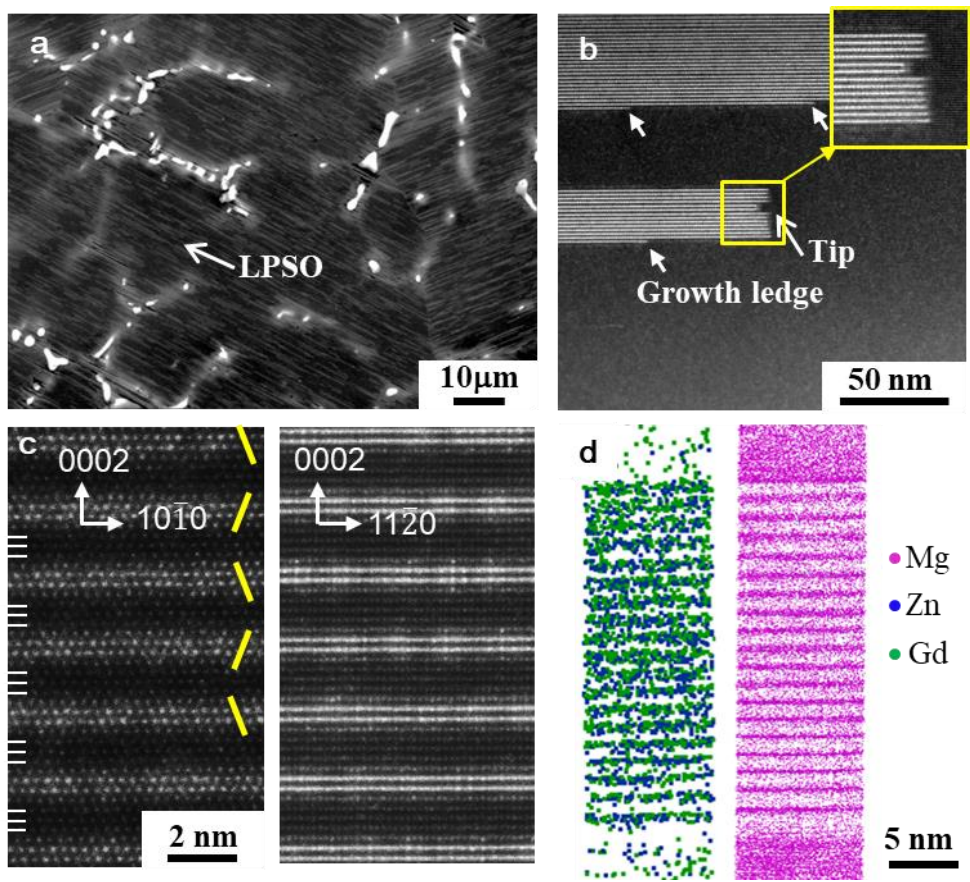

Figure 1 Microstructure of the sample aged at 500°C for 4h. a) SEM image. b) HAADF-STEM image. c) Atomic-resolved HADF-STEM image of 14H type LPSO structure at the zone axes of $[1\bar{2}10]_{hcp}$ and $[1\bar{1}00]_{hcp}$. d) The alternative distribution of Zn and Gd elements in 14H type LPSO structure measured by 3DAP.

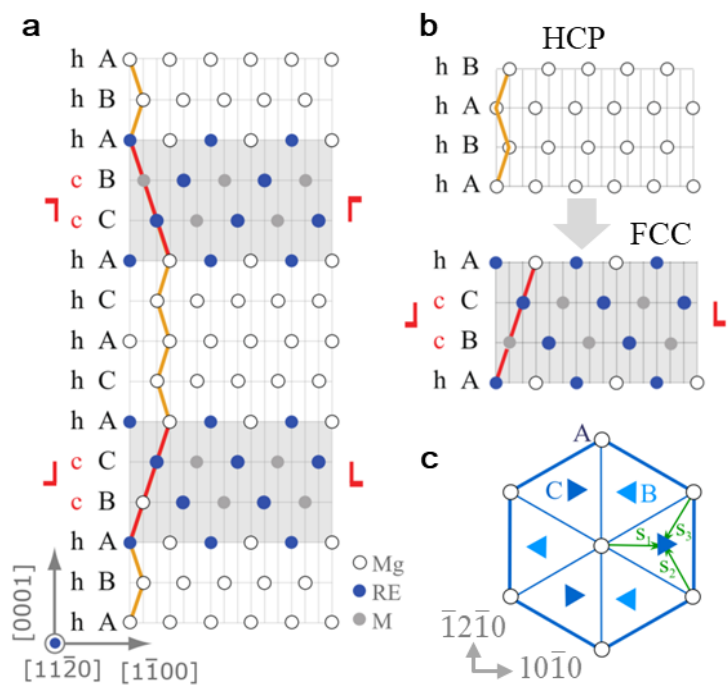

Figure 2 Schematic diagram of 14H type LPSO structure. a) $<11\bar{2}0>_{hcp}$ view of 14H type LPSO structure. b) The transformation from hcp structure to fcc structural unit by a shear process. c) Three possible shear directions for changing a stacking layer from A to B. (For interpretation of the references to color in this figure legend, the reader is referred to the web version of this article.)

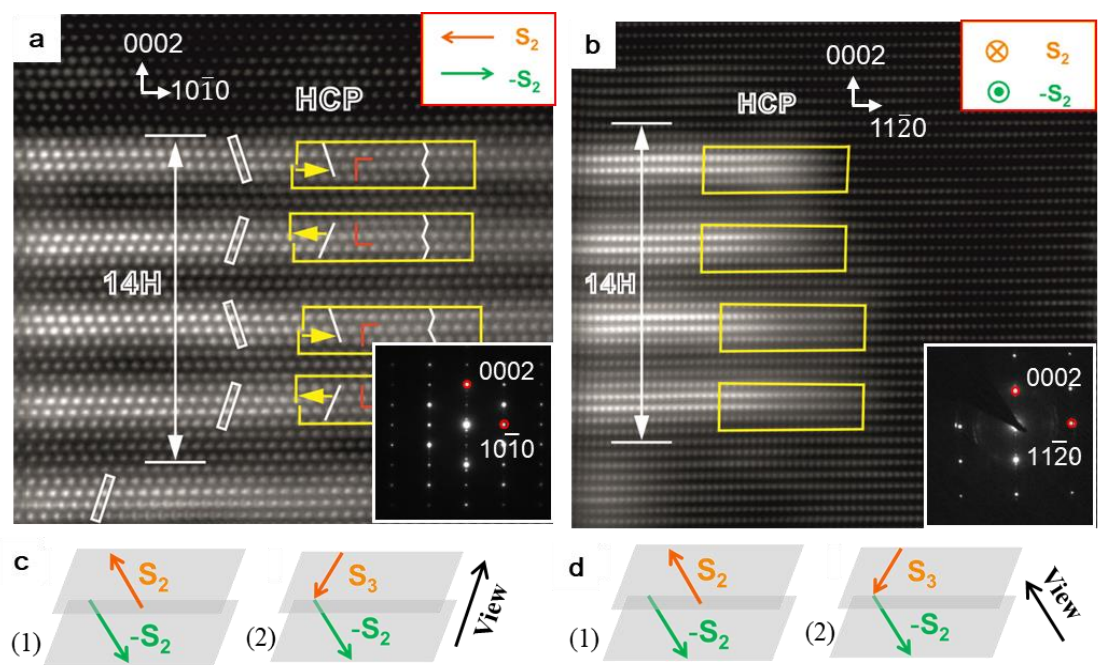

Figure 3 HAADF-STEM images of the transformation front between 14H type LPSO structure and Mg matrix. a) $[1\bar{2}10]_{hcp}$ zone axis. b) $[1\bar{1}00]_{hcp}$ zone axis. c) Two possible combinations of shear directions and the view direction for a). d) Two possible combinations of shear directions and the view direction for b). The insets in (a-b) are corresponding diffraction patterns.

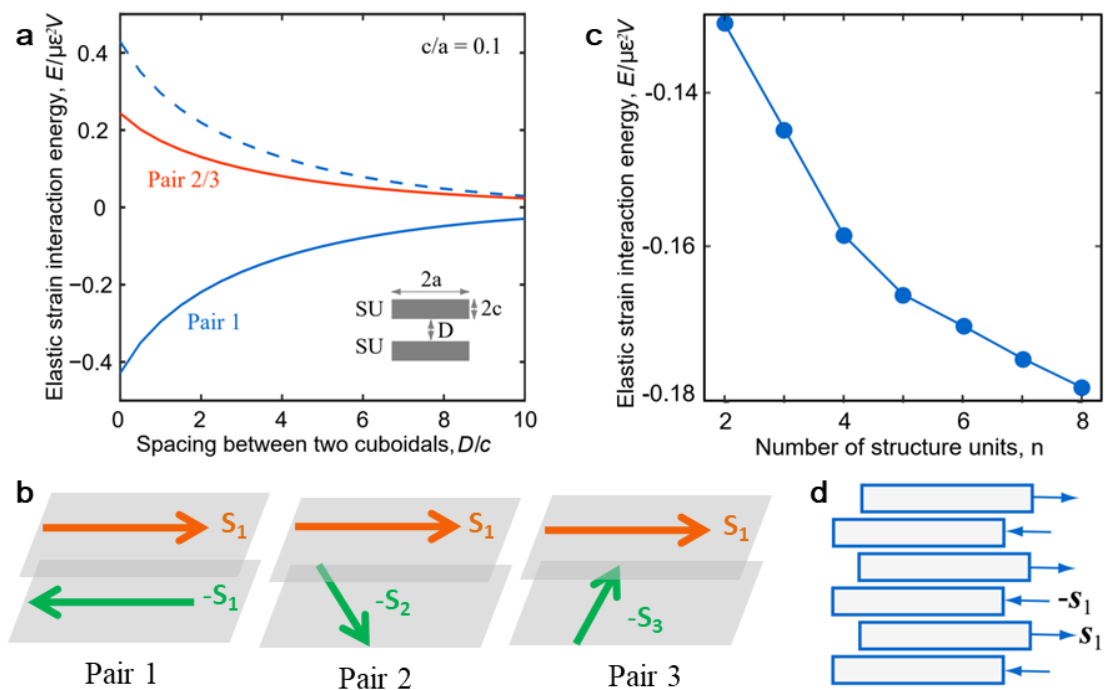

Figure 4 Elastic interaction energy for different combinations of shear directions. a) Variation of elastic interaction energy with the interspacing between two structural units for the configurations shown in b). The elastic interaction energy is scaled by $\mu\varepsilon^2 V$ where V is the volume of single unit, $\mu$ is the shear modulus, $\varepsilon$ is the shear strain and the negative dilatational strain is set to be -$\varepsilon$. b) Possible combinations of shears between neighbor structural units. c) The interaction energy between multiple structural units. d) The low energy configuration of shear directions.

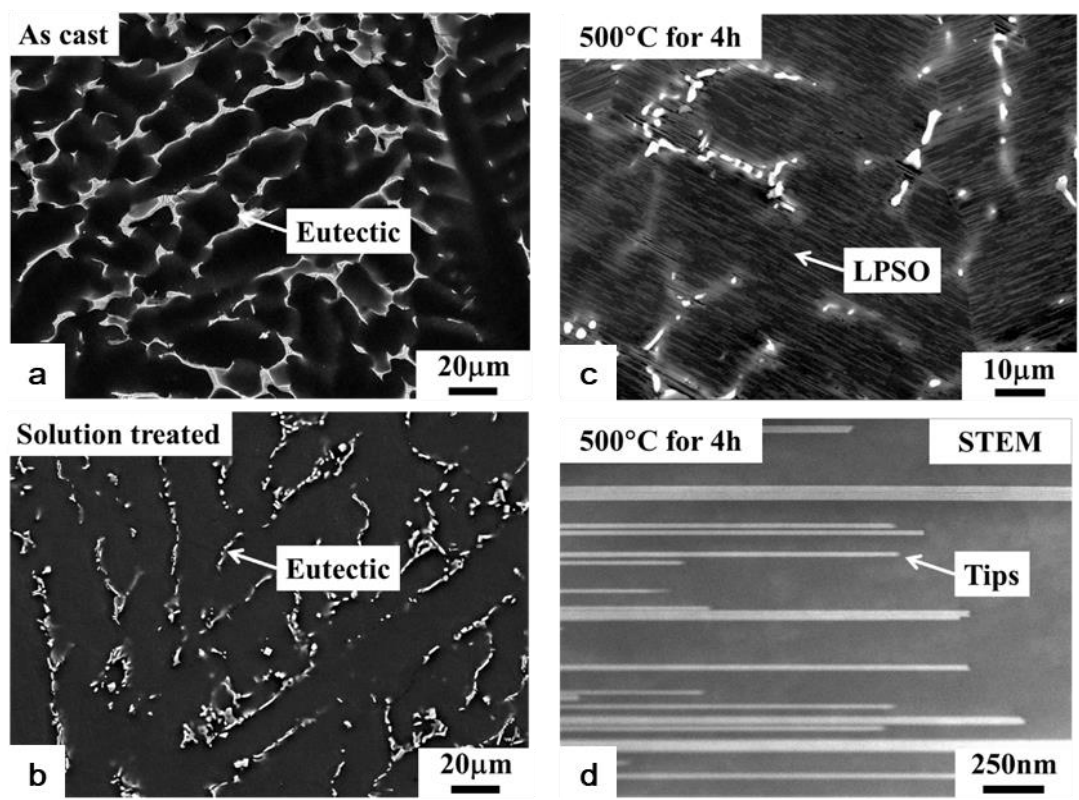

Figure S1 Microstructure of Mg-1Zn-2Gd alloy at various states. (a) SEM micrograph at as-casted state, (b) SEM micrograph at solution treated state, (c) SEM micrograph for aged at 500°C for 4h, (d) Low mag of HAADF-STEM image for (c).

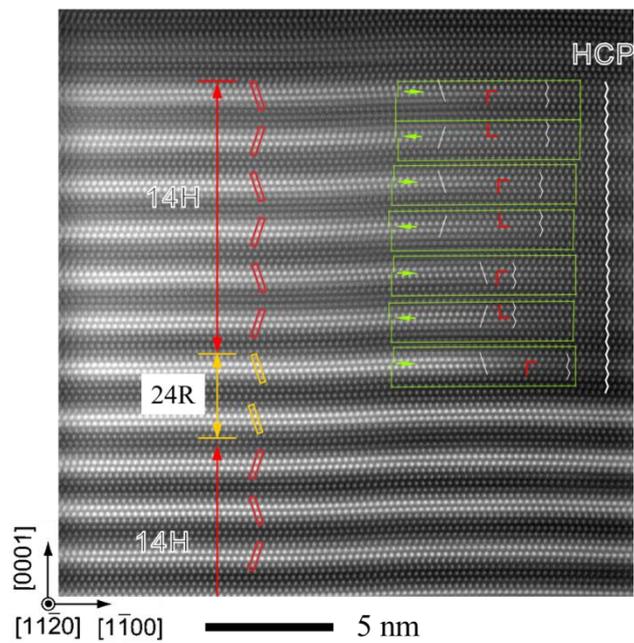

Figure S2 The Burgers circuits around the tips of SUs in the sample aged at 500°C for 4h.

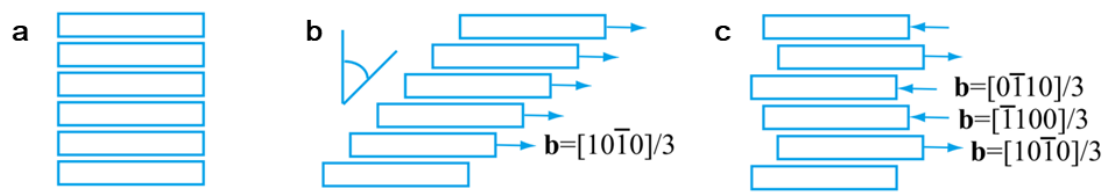

Figure S3 Schematic diagram of the transformation process for 18R type LPSO structure. (a) HCP matrix, (a) 18R by single partial, (b) self-accommodated by multiple partials.

**Graphic abstract**

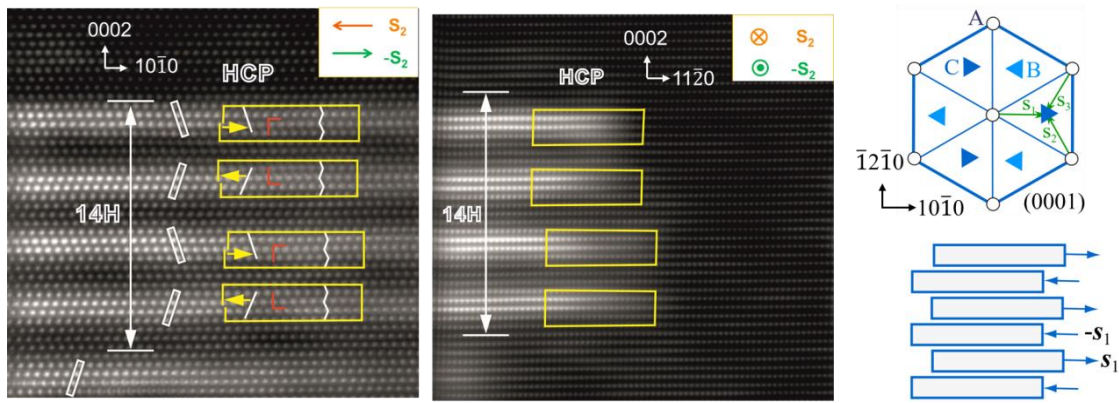